\documentclass{aip-cp}

\usepackage[numbers]{natbib}
\usepackage{rotating}
\usepackage{graphicx}

\usepackage{color}
\usepackage{caption}
\usepackage{subcaption}

\newcommand{\CHAIN}[1]{\mathbf{#1}}
\renewcommand{\P}{{\CHAIN{P}}}
\newcommand{\Q}{{\CHAIN{Q}}}

\begin{document}

\title{An Information Physics Derivation of Equations of Geodesic Form from the Influence Network}

\author[aff1]{James Lyons Walsh\corref{cor1}}
\author[aff1,aff2]{Kevin H. Knuth}
\eaddress{kknuth@albany.edu}

\affil[aff1]{Department of Physics, University at Albany (SUNY), 1400 Washington Avenue, Albany, NY 12222 USA}
\affil[aff2]{Department of Informatics, University at Albany (SUNY), 1400 Washington Avenue, Albany, NY 12222 USA}
\corresp[cor1]{
jlwalsh@albany.edu}




 \maketitle

\begin{abstract}
Information physics considers physical laws to result from the consistent quantification and processing of information about physical phenomena. In previous efforts, one of us (Knuth) has shown that a simple model of particles that directly influence one another results in a partially ordered set referred to as the influence network, from which emerge the Minkowski metric and Lorentz transformations of special relativity. Here, we extend earlier work on receipt of influence to the case of one particle influencing another, finding that this gives rise to equations of the form of geodesic equations from general relativity in 1+1 dimensions. Future work will test the equivalence of the current result to general relativity in 1+1 dimensions.
\end{abstract}


\section{Background}

Information physics \cite{Knuth:infophysics} \cite{Goyal:2012information} \cite{Caticha2012:entropic} contends that at least some of the laws of physics result from the consistent quantification and optimal information processing of information obtained by observers about the physical world.  Knuth and Bahreyni \cite{Knuth+Bahreyni:JMP2014} \cite{Bahreyni:Thesis} have relied on the concept of information physics to demonstrate that the consistent quantification of a partially ordered set of events with respect to an embedded observer, represented by a chain of events, results in the mathematics of special relativistic spacetime.  This was later extended by assuming that particles influence one another in a direct particle-particle interaction.  Each act of influence defines two events: the act of influence, which is associated with the influencing particle, and the act of being influenced, which is associated with the influenced particle.  The result is that particles are described by totally-ordered chains of influence events (emitted or received) that together form a network called the \textit{influence network} \cite{Knuth:FQXI2013} \cite{Knuth:Info-Based:2014}. The current work extends on past work \cite{Walsh+Knuth:acceleration} by examining the case of a particle chain receiving influence from another particle chain and showing that consistent quantification of the particle's behavior is consistent in form with general relativity in $1 + 1$ dimensions.

To illustrate the influence network, consider the case of the \textit{free particle}, which influences others but is not itself influenced. Events on a chain are quantified by mapping them to elements of a totally ordered set.  Without loss of generality, this can be accomplished by simply labeling the events along a chain with integers or real numbers so that if event $x$ precedes $y$ in the total order, denoted $x \leq y$ and read ``$y$ includes $x$,'' then they are associated with numbers $q(x)$ and $q(y)$ such that $q(x) \leq q(y)$, where in the context of real numbers the symbol $\leq$ means the usual less-than-or-equal-to.\footnote{For a more detailed description of the assumptions underlying the partially-ordered set of events, please see \cite{Knuth:Info-Based:2014}.} Since it has been previously demonstrated that there exists a unique description of events (up to scale) by embedded observers that results in the mathematics of special relativistic spacetime, one can talk about events in terms of the partially ordered set (poset), which is referred to as the \textit{poset picture}, or equivalently in terms of space and time, which is referred to as the \textit{spacetime picture}. Given these descriptions, one may consider the ordering of events as a causal order in a spacetime.

\begin{figure}
	\centering
	[a]
	\begin{minipage}{0.20\textwidth}
		\includegraphics[width=\textwidth]{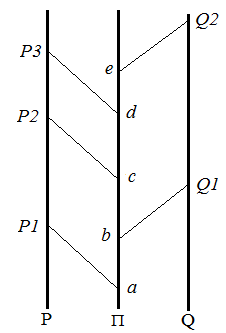}
	\end{minipage}
	[b]
	\begin{minipage}{0.295\textwidth}
		\includegraphics[width=\textwidth]{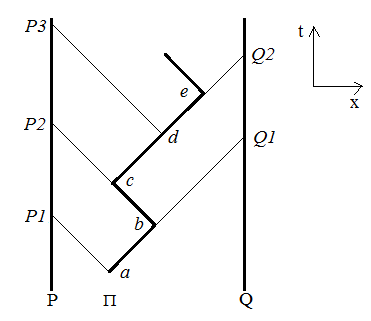}
	\end{minipage}
	\caption{\textbf{a:} In this Hasse diagram, the events where the free particle $\Pi$ influences the observer $\P$ are indicated by the elements $a$, $c$, and $d$ along the particle chain $\Pi$.  The events $P1$, $P2$, and $P3$ along the chain $\P$ represent the corresponding acts of observer $\P$ being influenced by $\Pi$. Likewise, the elements $b$ and $e$ along the chain $\Pi$ indicate the acts of particle $\Pi$ influencing observer $\Q$, which result in the events representing $\Q$ being influenced by $\Pi$: $Q1$ and $Q2$, respectively. Note that in a Hasse diagram, the lengths of the lines are irrelevant, since the only thing that is important is the connectivity of the diagram. \textbf{b:} In the spacetime picture, the result is that the particle $\Pi$ is described by the observers $\P$ and $\Q$ as moving in spacetime in a way that corresponds to the Hasse diagram.}
	\label{free_particle_Hasse_diagram}
\end{figure}

One important property that observer chains are defined to possess is  \textit{coordination}, which is the existence of a bijectivity between their events and the condition of equality of the lengths of corresponding intervals on the chains. In the spacetime picture, this corresponds to the fact that the observers whose clocks agree with one another define an inertial frame. Observers bear a particular relation to particles as well as to each other, which is based on the concept of \textit{projection}.

The \textit{forward projection} of an event onto a chain (if such a projection exists) maps the event to the least event on the chain that includes it. The \textit{backward projection}, or back projection, of an event onto a chain maps the event to the greatest event on the chain that it includes.
Figure \ref{free_particle_Hasse_diagram}a
depicts the Hasse diagram of the free particle chain $\Pi$ influencing observers $\P$ and $\Q$, with the influences represented by the angled fine lines. Projection depends on the influence connectivity, so events $a$, $c$, and $d$ forward project to $\P$ at $P1$, $P2$, and $P3$, respectively, and events $b$ and $e$ forward project to $\Q$ at $Q1$ and $Q2$, respectively. 
To project an event that is not one of influencing the chain being projected to, such as event $a$ to observer $\Q$, we note that projection is transitive via influence, and forward projection is to the least event that includes the projected event. The event of reception of the next influence from $\Pi$ to the chain is the least event on the chain that includes the event on $\Pi$ to be projected. Therefore, event $a$ forward projects to $\Q$ at $Q1$; event $b$ forward projects to $\P$ at $P2$; and both events $c$ and $d$ forward project to $\Q$ at $Q2$.
These considerations give rise to the diagram representing the spacetime picture in Figure \ref{free_particle_Hasse_diagram}b.

Since the free particle is not influenced (or equivalently, does not receive influence), back projection of its events must be performed by assuming that the particle chain's events possess the property of \textit{collinearity} with their projections onto the observer chains, which is the relation between the particle and the observers that defines the 1 + 1 dimensional subspace that the observer chains induce in the poset
\cite{Knuth+Bahreyni:JMP2014} \cite{Bahreyni:Thesis}. Collinearity of an event with its projections onto chains is the property that projection to one of two chains can be found by projecting to the other chain and then projecting to the one. For example, the event $a$ can be back projected to $\Q$ by first forward projecting to $P1$.  Likewise, since event $b$ forward projects to $\P$ at $P2$, it back projects to $\Q$ at the same event that event $c$ does. The mechanism of back projection of an event that is not one of reception of influence will be described in future work.

Observers quantify events and intervals bounded by events by projecting them onto their chains.
A sufficiently small interval on the particle chain $\Pi$, representing a differential interval, projects to differential intervals $dp$ and $dq$ on the chains $\P$ and $\Q$.
These intervals are differential not through a limiting process but by being small in comparison to the smallest amount that is actually being measured by a macroscopic agent. The observer chains are sensitive to individual events, and the derivation below will be based on the full sensitivity of the observers until calculus is employed. It will be seen that there are at least three length scales involved: the length of a particle chain between successive emissions of influence, $\lambda_e$; the length, $\lambda_r$, of the particle chain between receptions of influence that are successive in the sense that there are only events of emission between them and no intervening receptions; and the smallest length measurable by the macroscopic agent, $\lambda_m$. When $\lambda_r \ll \lambda_m$, $\lambda_r$ can be treated as a differential amount, and calculus can be used based on a definition of continuity akin to the continuum hypothesis in fluid dynamics.

We now consider a special case where we have a particle that makes a transition, via $N$ events, from an initial state to a final state.
Lengths of subintervals between successive receptions of influence from the particle on the observer chains have been shown in \cite{Knuth+Bahreyni:JMP2014} and \cite{Bahreyni:Thesis} to be inversely related for our choice of scale.\footnote{Note that in what follows, $k$ represents a different quantity than it did in \cite{Knuth+Bahreyni:JMP2014}. The variable $k$ as used here is $\sqrt{\frac{m}{n}}$ in \cite{Knuth+Bahreyni:JMP2014} in units determined by the choice of scale.} With $k_{p}$ being the length of a subinterval on $\P$, $k_{q}$ being the length of a subinterval on $\Q$, and with $N_{p}$ and $N_{q}$ being the number of subintervals on $\P$ and $\Q$, respectively, the differential intervals $dp$ and $dq$ are given by \cite{Knuth+Bahreyni:JMP2014} \cite{Walsh+Knuth:acceleration}
\begin{equation} \label{eq:observer_intervals}
dp = N_{p} k_{p} = N_{p} k \qquad \qquad dq = N_{q} k_{q} = N_{q} \frac{1}{k}.
\end{equation}
Likewise, it is shown by means of linearly-related chains that the unique consistent scalar measure, $d \tau$, of a differential interval on a particle chain is given by \cite{Knuth+Bahreyni:JMP2014} \cite{Walsh+Knuth:acceleration}
\begin{equation} \label{eq:interval}
d \tau^{2} = dp dq = N_{p} k N_{q} \frac{1}{k} = N_{p} N_{q},
\end{equation}
which corresponds in the spacetime picture to the square of the proper time. By means of a simple change of variables suggested by the symmetric and antisymmetric decomposition, the quantities describing the time and space components of an interval in the spacetime picture are
\begin{equation} \label{eq:dt_and_dx}
dt = \frac{dp + dq}{2} \qquad \qquad dx = \frac{dp - dq}{2},
\end{equation}
which can be seen to obey $d \tau^{2} = dpdq = dt^{2} - dx^{2}$.

The result obtained below depends on imposing consistency between the length of a particle chain interval given by $d \tau^2 = dpdq$ and that given by the sum of the lengths of particle chain subintervals bounded by successive events, which we call \textit{atomic intervals}. As long as they are bounded by events having unequal valuations, atomic intervals on a particle chain are of equal length $\delta \tau$, since the valuations of events are assigned with uniform increment. There exists an observer pair for which $k = \frac{1}{k} = 1$. From \cite{Knuth+Bahreyni:JMP2014}, the usual definition of velocity as the ratio of the differential increment in $x$ to that in $t$ is accompanied by a second definition that gives rise to relations analogous to the Lorentz transformations:
\begin{equation} \label{eq:v(k)}
v = \frac{dx}{dt} =  \frac{k - \frac{1}{k}}{k + \frac{1}{k}}.
\end{equation}
Thus, the observer pair for which $k = 1$ defines the rest frame. We have chosen our scale so that the length along the particle chain is the time in the rest frame, which makes $\delta \tau = \frac{1}{2}$, by (\ref{eq:observer_intervals}) and (\ref{eq:dt_and_dx}). Note that since the value of the interval scalar given by (\ref{eq:interval}) depends only on numbers of events, it is naturally Lorentz invariant, dependence on velocity through $k$ having canceled.

\section{The Influenced Particle}

\begin{figure}
\centering
	[a]
	\begin{minipage}{0.25\textwidth}
		\includegraphics[width=\textwidth]{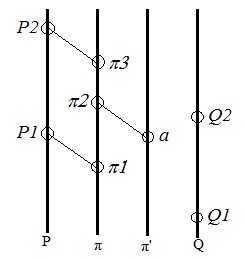}
	\end{minipage}
	[b]
	\begin{minipage}{0.26\textwidth}
		\includegraphics[width=\textwidth]{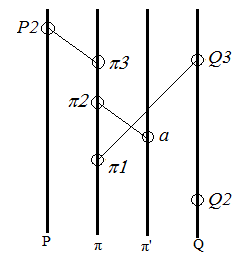}
	\end{minipage}
	\caption{This figure illustrates two situations each involving several influence events, discussed in the text.  Note that not all influence events are illustrated.  The assumption of collinearity is consistent with the situation illustrated in \textbf{a} and is inconsistent with the situation illustrated in \textbf{b}.}
	\label{Influence Hasse Diagram}
\end{figure}

In this section, we explore situations in which a particle not only influences but also is influenced.
Figure \ref{Influence Hasse Diagram}a illustrates a situation involving two particles and several influence events.  Note that not all influence events are illustrated.  Events $\pi 1$ and $\pi 3$ are acts of particle $\pi$ influencing observer $\P$ at events $P1$ and $P2$, respectively, where $P2 > P1$. By coordination, there exist events $Q1$ and $Q2$, where $Q2 > Q1$, such that events 
$P1$ and $P2$ back project to observer $\Q$ at events $Q1$ and $Q2$, respectively.  Event $a$ denotes the act of particle $\pi '$ influencing particle $\pi$, whereas the reception of influence by $\pi$ is denoted by event $\pi 2$. Since both events $\pi2$ and $\pi3$ forward project to $P2$, and $P2$ back projects to $Q2$, the assumption of collinearity implies that events $\pi2$, $\pi3$, and $a$ must all back project to $Q2$.  Similarly, the assumption of collinearity implies that event $\pi1$ must back project to $Q1$.

One can show that when a particle is influenced, it is not guaranteed that the ordering of events along the particle chain will be consistent with the assumption of collinearity. That is, the assumption of collinearity puts constraints on the order in which events can occur along the particle chain.  For example, in Figure \ref{Influence Hasse Diagram}a, event $a$ forward projects to $\P$ at $P2$. Thus $P2$ includes event $a$.  By the definition of back projection, there exists an event $b \in \pi '$ (not shown) at which $P2$ back projects to $\pi '$. It may be that event $b = a$, but this has not been demonstrated. In the case where $b = a$, it would be true that $\pi$ was collinear with $\P$ and $\pi '$.  However, if $b \neq a$, then collinearity would be violated.

Another example is illustrated in Figure \ref{Influence Hasse Diagram}b where event $\pi 1$ denotes the act of $\pi$ influencing $\Q$ at event $Q3$. By the definition of back projection, there exists an event $c \in \Q$ (not shown) at which $\pi 1$ back projects to $\Q$. We again assume that there exists an event $Q2 \in \Q$, such that $P2$ back projects onto $\Q$ at event $Q2$.  In the case where $c = Q2$, we have that the situation is consistent with collinearity, since $\pi 1$ forward projects onto $\P$ at $P2$, and both $P2$ and $\pi 1$ back project onto $\Q$ at $c = Q2$.  However, if $c \neq Q2$, then collinearity is violated.

Collinearity requires that we could find the back projection of $\pi 1$ to $\pi '$ by forward projecting to $\P$ at $P2$ and then back projecting to $\pi '$ at event $b = a$. This means that $\pi 1$ would include event $a$. Since event $a$ forward projects to $\pi 2$, by the definition of forward projection we have that $\pi 2$ must be the least event on the chain $\pi$ that includes $a$. However, $\pi 1 < \pi 2$, and since we have shown that collinearity and back projection imply that $\pi 1$ includes $a$, we have a contradiction since $\pi 1$ and not $\pi 2$ must be the least event on $\pi$ that includes $a$.  This contradiction implies that the assumption, which is that of collinearity, cannot hold in this situation. As a result, we conclude that the assumption of collinearity requires that an event of receiving influence from one side cannot be immediately preceded by an event of influencing to the same side.  That is, such a situation would violate collinearity.

We return to Figure \ref{Influence Hasse Diagram}a, which we shall refer to henceforth. The successive events $\pi 2$ and $\pi 3$ both forward project to event $P2$ on $\P$.  Furthermore, since neither of the successive events $\pi 2$ or $\pi 3$ represents the particle influencing $\Q$, they must both forward project to the same event $Q3$ (which is not pictured in Figure \ref{Influence Hasse Diagram}a) such that $Q3 > Q2$. Since $\pi 2$ and $\pi 3$ both forward project to the same event on each observer ($P2$ on $\P$ and $Q3$ on $\Q$), the atomic interval $[\pi 2 ,\pi 3]$ on the particle chain is measured by the observers as having zero length.
Furthermore, events $\pi 1$ and $\pi 2$ both forward project to event $Q3$, so $\delta q = 0$ for atomic interval $[\pi 1 ,\pi 2]$. The question is, what is $\delta p$ for $[\pi 1, \pi 2]$?

There are two possibilities.
One is that $\delta p = k$, such that the event $\pi 2$ has no effect, and the particle remains free. The other possibility is that $\pi 2$ has some effect, and $\delta p \neq k$. Because $\pi 2$ and $\pi 3$ project to the same event on $\P$, all that $\P$ sees is that there are two events, $P1$ and $P2$, of receiving influence from the particle with a new length $k'$ for the interval between them. To keep the scale that we've selected fixed, $k_q$ must remain equal to the reciprocal of $k_p$, therefore becoming $\frac{1}{k'}$. The atomic interval lengths have changed, but we don't know how yet.

Because there is a corresponding change in $\frac{1}{k}$ for each change in $k$, the observers can associate the events at which these changes take place on their chains with particular spacetime locations of the particle.
The uncertainty in this association is a topic for future work. In particular, the observers know that the event immediately preceding the change in $k$ on $\P$ and the event immediately preceding the change in $\frac{1}{k}$ on $\Q$ both back project to $\pi 1$, giving its location. In addition, it will be shown that $k$ changes in different ways for reception from the $\Q$ and $\P$ sides, so the observers also know that the event the location of which they've found
is an event of $\pi$ influencing $\P$, giving them the spacetime location of $\pi 2$ and $\pi 3$, both of which have the same location. 
Location of the particle can be defined only when there is consistency between the length along the particle chain given by (\ref{eq:interval}) and the sum of the lengths of the atomic intervals. That the location is known implies that the consistency requirement holds.

The consistency requirement takes the form of 
\begin{equation} \label{consistency_requirement}
 dpdq = \bigg( \frac{N}{2} \bigg)^2,
\end{equation}
where $N$ is the number of events of influencing, which equals the number of atomic intervals of nonzero length, $\delta \tau$ being equal to $\frac{1}{2}$ for each of these atomic intervals. This implies $N_p = N_q = \frac{N}{2}$ from (\ref{eq:interval}) and the fact that $N_q = N - N_p$. Here, $N_p$, $N_q$, and $N$ are numbers of influences sent by the particle, starting with the event at which its position could last be inferred and ending with the event 
just before $\pi 1$. This also satisfies the requirement that $k$ and the velocity correspond on that interval, in accordance with (\ref{eq:v(k)}).

Additionally required is the consistency expressed in (\ref{consistency_requirement}) for the union of the interval containing the $N$ atomic intervals with $[\pi 1 ,\pi 2]$, since the observers have inferred the location of $\pi 2$. The additional atomic interval adds $\frac{1}{2}$ to $d \tau$, replacing $N$ with $N+1$ in (\ref{consistency_requirement}).
Note that events, such as $\pi 2$ and $\pi 3$, that
the observers cannot distinguish are assigned the same valuation. This gives that $dp \rightarrow (\frac{N+1}{N})^2 dp$ from the fact that $\pi 1$ and $\pi 2$ forward project to the same event on $\Q$, so that $dq$ cannot change.
However, we have another way to compute the length of $dp$, which results from considering $N_p$ intervals of length $k$ plus an additional interval of length $k_1$.
This new length, $k_1$, of the projection of the interval $[\pi 1, \pi 2]$ onto $\P$ can be found by setting these two expressions for $dp$ equal to one another so that
\begin{equation}
N_p k + k_1 = \bigg( \frac{N+1}{N} \bigg)^2 N_p k
\end{equation}
is solved to give
\begin{equation}
k_1 = \frac{N+\frac{1}{2}}{N} k.
\end{equation}
Using (\ref{eq:dt_and_dx}) and (\ref{eq:v(k)}), the new velocity is
\begin{equation}
v' = \frac{dx'}{dt'} = \frac{\frac{N}{2} k + \frac{N+\frac{1}{2}}{N} k - \frac{N}{2} \frac{1}{k}}{\frac{N}{2} k + \frac{N+\frac{1}{2}}{N} k + \frac{N}{2} \frac{1}{k}}
= \frac{\frac{N+1}{N} k - \frac{N}{N+1} \frac{1}{k}}{\frac{N+1}{N} k + \frac{N}{N+1} \frac{1}{k}} = \frac{k' - \frac{1}{k'}}{k' + \frac{1}{k'}},
\end{equation}
which gives that $k' = \frac{N+1}{N} k$.

To summarize, if $k$ changes, the observers are able to localize the particle in spacetime. Since location is defined only when consistency requirement (\ref{consistency_requirement}) holds, this requirement is known to apply. Applying (\ref{consistency_requirement}) gives the length of $\delta p$ corresponding to $[ \pi 1, \pi2]$. This, in turn, gives the new velocity through the first expression in (\ref{eq:v(k)}). The additional consistency requirement embodied in the latter equality of (\ref{eq:v(k)}) gives $k'$, the new value of $k$. The consequences of the amount by which $k$ changes will be explored below. Establishing the conditions under which $k$ does and does not change upon reception will be a subject of future work. In what follows, the term ``reception'' denotes a reception associated with change in $k$.

This has given $dp$ for a particle chain interval bounded by events of receiving influence that are successive in the sense that there are only events of emission between them and no intervening receptions, in which the second reception is from the $\Q$ side. It has also given $k'$, the value of $k$ following the second event of being influenced. We can use the same arguments to find that upon an event of reception from the $\P$ side, $dq \rightarrow (\frac{N+1}{N})^2 dq$ for the interval between successive events of being influenced, and $k' = \frac{N}{N+1} k$. For simplicity of calculation and in particular to use a single value of $k$ for all the atomic intervals in an interval bounded by successive events of being influenced, we can write effective values of $N_p$, $N_q$, and $k$ as follows.
\begin{equation} \label{eq:qbar_changes}
N_p = N_q = \frac{N}{2} + \frac{1}{2} \qquad \qquad k \rightarrow \frac{N+1}{N} k \qquad \qquad \mathrm{(reception} \, \mathrm{from} \, \Q \, \mathrm{side)}
\end{equation}
\begin{equation} \label{eq:pbar_changes}
N_p = N_q = \frac{N}{2} + \frac{1}{2} \qquad \qquad k \rightarrow \frac{N}{N+1} k \qquad \qquad \mathrm{(reception} \, \mathrm{from} \, \P \, \mathrm{side)}
\end{equation}
That is, the particle chain interval bounded by receptions and ending in $\pi2$ is quantified the same as it would be with $k$ replaced by $k'$ and the atomic interval $[\pi 1 ,\pi 2]$ replaced by halves of the atomic intervals that would result from the particle influencing each observer once.

\section{Geodesic Equations}

The next step is to write differentials. 
Continuity arises as noted above from the fact that these differentials are much smaller than the smallest measurable amount. These lengths are not taken to zero but rather are treated in a way akin to the continuum hypothesis in fluid dynamics. Since the $k$ values change with the receipt of each influence, it is convenient to choose the differential amounts to correspond to intervals between successive events of receiving influence. With $r_{\bar{q}} = \frac{1}{N_{\mathbf{Q}}}$ being the mean rate per event at which the particle is influenced from the $\Q$ side, where $N_{\mathbf{Q}}$ is $N+1$ in (\ref{eq:qbar_changes}), 
and $r_{\bar{p}} = \frac{1}{N_{\mathbf{P}}}$ being the mean rate at which the particle is influenced from the $\P$ side, $N_{\mathbf{P}}$ similarly being $N+1$ in (\ref{eq:pbar_changes}), 
the number of events of influencing the observers in the interval between successive reception events is on average the reciprocal of the total rate:
\begin{equation} \label{eq:average_N}
N' = N + 1 = \frac{1}{r_{\bar{q}} + r_{\bar{p}}},
\end{equation}
where we define
\begin{equation}
\tilde{r} \, \dot{=} \, r_{\bar{q}} + r_{\bar{p}}.
\end{equation}
The number $N$ does not include the last event of influencing the observers  before reception. The probabilities that the interval between receptions ends in a reception from a given side are
\begin{equation} \label{eq:probabilities}
Pr(\textrm{reception from $\Q$ side}) = \frac{r_{\bar{q}}}{\tilde{r}} \qquad Pr(\textrm{reception from $\P$ side}) = \frac{r_{\bar{p}}}{\tilde{r}}.
\end{equation}
Here we consider the case in which these probabilities do not depend on the length of the interval.

In this case, we can find the average increments in $p$ and $q$ due to one interval between receptions from equation (\ref{eq:observer_intervals}) in terms of effective values as
\begin{equation}
dp = \langle N_p k' \rangle = \langle N_p \rangle \langle k' \rangle = \frac{1}{2 \tilde{r}} \langle k' \rangle \qquad \qquad dq = \bigg\langle N_q \frac{1}{k'} \bigg\rangle = \langle N_q \rangle \bigg\langle \frac{1}{k'} \bigg\rangle = \frac{1}{2 \tilde{r}} \bigg\langle \frac{1}{k'} \bigg\rangle.
\end{equation}
Using the fact from (\ref{eq:average_N}) that $N = \tilde{r}^{-1} - 1$ and equations (\ref{eq:qbar_changes}), (\ref{eq:pbar_changes}), and (\ref{eq:probabilities}), this yields the following.\footnote{While to lowest order, the first expression corresponds to the insertion of a single increment $k$ in $dp$ upon the receipt of an influence from the $\Q$ side when $r_{\bar{p}} = 0$, as was discussed in \cite{Walsh+Knuth:acceleration}, the approach in this derivation is very different.}
\begin{equation}
dp = \frac{1}{2 \tilde{r}} \Biggl[ \frac{r_{\bar{q}}}{\tilde{r}} \frac{\tilde{r}^{-1}}{\tilde{r}^{-1} - 1} + \frac{r_{\bar{p}}}{\tilde{r}} \frac{\tilde{r}^{-1} - 1}{\tilde{r}^{-1}} \Biggr] k = \frac{1}{2 \tilde{r} (1-\tilde{r})} (1 - 2 r_{\bar{p}} + r_{\bar{p}} \tilde{r}) k
\end{equation}
\begin{equation}
dq = \frac{1}{2 \tilde{r}} \Biggl[ \frac{r_{\bar{q}}}{\tilde{r}} \frac{\tilde{r}^{-1} - 1}{\tilde{r}^{-1}} +  \frac{r_{\bar{p}}}{\tilde{r}} \frac{\tilde{r}^{-1}}{\tilde{r}^{-1} - 1} \Biggr] \frac{1}{k} = \frac{1}{2 \tilde{r} (1 - \tilde{r})} (1 - 2 r_{\bar{q}} + r_{\bar{q}} \tilde{r}) \frac{1}{k}
\end{equation}
Here, $k$ is the effective value on the previous interval between receipts of  influence. The change in the differential increments in time and space, $dt$ and $dx$, from one interval between receptions to the next are given by (\ref{eq:dt_and_dx}) as\footnote{Equation (\ref{eq:ddt_and_ddx}) is two equations, one for ddt and one for ddx.}
\begin{equation} \label{eq:ddt_and_ddx}
\begin{array}{c}ddt \\ ddx \end{array} = \frac{1}{4} \Biggl[ \frac{1}{\tilde{r}(1-\tilde{r})} \Biggl( (1 - 2 r_{\bar{p}} + r_{\bar{p}} \tilde{r}) k \pm (1 -2 r_{\bar{q}} + r_{\bar{q}} \tilde{r}) \frac{1}{k} \Biggr) -  \frac{1}{\tilde{r}_{0}} \Biggl(k \pm \frac{1}{k} \Biggr) \Biggr].
\end{equation}
Note that $k$ was the known value for the initial increments, which took into account the influence received on that interval. We also have from the fact that function values in differentials are their initial values that
\begin{equation} \label{eq:dtau,dt,dx}
d \tau = \frac{1}{2 \tilde{r}_{0}} \qquad \textrm{and} \qquad \begin{array}{c} dt \\ dx
\end{array} = \frac{d \tau}{2} \Biggl(k \pm \frac{1}{k} \Biggr).
\end{equation}
Because we are considering the case in which continuity applies, we can approximate $\tilde{r} = \tilde{r}_{0} + d \tilde{r} \approx \tilde{r}_{0}$ and similarly for $r_{\bar{p}}$ and $r_{\bar{q}}$. With this, the definition $r \dot{=} r_{\bar{q}} - r_{\bar{p}}$, and the decomposition of the factors multiplying $k$ and $\frac{1}{k}$ into their average and half-difference, we can rewrite (\ref{eq:ddt_and_ddx}) as
\begin{equation} \label{eq:simplified_ddt_and_ddx}
\begin{array}{c}
ddt \\ ddx
\end{array} =  \frac{\tilde{r}}{8(1-\tilde{r})} \Biggl(k \pm \frac{1}{k} \Biggr) + \frac{r}{4 \tilde{r}} \Biggl(1 +\frac{\tilde{r}}{2 (1-\tilde{r})} \Biggr) \Biggl( k \mp \frac{1}{k} \Biggr) .
\end{equation}
Divide both sides of (\ref{eq:simplified_ddt_and_ddx}) by $d \tau^{2}$, and rearrange to obtain from $ddt$
\begin{equation}
\frac{d^{2} t}{d \tau^{2}} = \frac{\tilde{r}^{2}}{2(1 - \tilde{r}) d \tau} \frac{dt}{d \tau} + \frac{r}{d\tau} \Biggl( 1 + \frac{\tilde{r}}{2 (1 - \tilde{r})} \Biggr) \frac{dx}{d\tau}.
\end{equation}
It can be seen from the presence of $d \tau$ in their denominators that the coefficients of $\frac{dt}{d \tau}$ and $\frac{dx}{d \tau}$ are rates. Therefore, each coefficient can be written as the total derivative of some quantity.
\begin{equation}
\frac{d\tilde{R}}{d \tau} \dot{=} \frac{\tilde{r}^{2}}{2 (1 - \tilde{r}) d \tau} 
\qquad \qquad \frac{dR}{d \tau} \dot{=} \frac{r}{d \tau} \Biggl( 1 + \frac{\tilde{r}}{2 (1 - \tilde{r})} \Biggr) 
\end{equation}
This gives
\begin{equation} \label{eq:geodesic_equation_for_t}
\frac{d^{2}t}{d \tau^{2}} = \frac{d \tilde{R}}{d \tau} \frac{dt}{d \tau} + \frac{dR}{d \tau} \frac{dx}{d \tau} = \frac{\partial \tilde{R}}{\partial t} \Biggl( \frac{dt}{d \tau} \Biggr)^{2} + \Biggl ( \frac{\partial \tilde{R}}{\partial x} + \frac{\partial R}{\partial t} \Biggr) \frac{dt}{d \tau} \frac{dx}{d \tau} + \frac{\partial R}{\partial x} \Biggl( \frac{dx}{d \tau} \Biggr)^{2},
\end{equation}
which is of the form of a geodesic equation for $t$, with the partial derivatives of $\tilde{R}$ and $R$ related to the Christoffel symbols. Following the same steps gives from $ddx$
\begin{equation} \label{eq:geodesic_equation_for_x}
\frac{d^{2}x}{d \tau^{2}} = \frac{d \tilde{R}}{d \tau} \frac{dx}{d \tau} + \frac{dR}{d \tau} \frac{dt}{d \tau} = \frac{\partial R}{\partial t} \Biggl( \frac{dt}{d \tau} \Biggr)^{2} + \Biggl ( \frac{\partial \tilde{R}}{\partial t} + \frac{\partial R}{\partial x} \Biggr) \frac{dt}{d \tau} \frac{dx}{d \tau} + \frac{\partial \tilde{R}}{\partial x} \Biggl( \frac{dx}{d \tau} \Biggr)^{2}.
\end{equation}

The equations of geodesic form, (\ref{eq:geodesic_equation_for_t}) and (\ref{eq:geodesic_equation_for_x}), show that the Christoffel symbols obey the coordinate conditions $2 \Gamma_{01}^{0} = \Gamma_{00}^{1} + \Gamma_{11}^{1}$ and $2 \Gamma_{01}^{1} = \Gamma_{00}^{0} + \Gamma_{11}^{0}$. Thus, the theory picks out a coordinate system. This will be studied in future work.

As a check on the results, consider a particle being influenced at a small, constant rate from the $\Q$ side: $\tilde{r} = r = r_{\bar{q}} = $ constant. Dropping factors higher than first order in the first equalities of (\ref{eq:geodesic_equation_for_t}) and (\ref{eq:geodesic_equation_for_x}) yields
\begin{equation}
\frac{d^{2}t}{d \tau^{2}} = \frac{r_{\bar{q}}}{d\tau} \frac{dx}{d \tau} \qquad \qquad \qquad \frac{d^{2}x}{d \tau^{2}} = \frac{r_{\bar{q}}}{d\tau} \frac{dt}{d \tau}.
\end{equation}
Note that since the rate of influence is constant, $d \tau$ is a constant as well. Solutions are
\begin{equation}
t = C_{1} \sinh \Biggl( \frac{r_{\bar{q}}}{d \tau} \tau + \phi_{0} \Biggr) + C_{2} \qquad \qquad \qquad x = C_{1} \cosh \Biggl( \frac{r_{\bar{q}}}{d \tau} \tau + \phi_{0} \Biggr) + C_{3}
\end{equation}
for initial rapidity $\phi_{0}$, where the $C$ values are constants. This is the result from special relativity for constant acceleration, with $\frac{r_{\bar{q}}}{d \tau} = 2 r_{\bar{q}}^2$ 
being the acceleration in the momentarily co-moving reference frame.\footnote{This result differs by a factor of 2 from that in \cite{Walsh+Knuth:acceleration}.}

As a further check on the result, it can be verified that the interval $d \tau$ is a maximum, as expected for a geodesic in general relativity. 
Using the effective values for convenience, we have from (\ref{eq:interval}) that $d \tau^2 = N_p N_q =  N_{p} (N' - N_{P})$, which has vanishing first derivative with respect to $N_{p}$ and negative second derivative when $N_{p} = \frac{N'}{2}$, where $N' = N_p + N_q = N + 1$. This is precisely the case, as shown in (\ref{eq:qbar_changes}) and (\ref{eq:pbar_changes}).
It will be a subject of future work to derive the metric.

\section{Conclusions}

The influence network, which has been shown to give rise to the physics of flat spacetime \cite{Knuth+Bahreyni:JMP2014}, here results in equations of the form of geodesic equations from general relativity when a particle is influenced by another. After the rates of influence reception have been found in terms of the mass distribution, it may result that in the influence network, a chain of events, some of which are receptions of influence, is quantified by macroscopic agents with limited measurement precision as following a geodesic path of the general relativistic free particle in spacetime. 
Unpublished work shows that time dilation that goes as the square of the rate at which influence is received results. 
The calculation of the rates of reception will be pursued, as will the extension of this work to $3 + 1$ dimensions by means of new results from one of us (Knuth).

\section{Acknowledgements}
We wish to thank Newshaw Bahreyni, Ariel Caticha, Seth Chaiken, Philip Goyal, Keith Earle, Oleg Lunin, Anthony Garrett, and John Skilling for interesting discussions and helpful questions and comments. We also thank two anonymous reviewers for comments.

\bibliographystyle{aipnum-cp}
\bibliography{knuth}

\end{document}